\documentclass[preprint,12pt]{elsarticle}
\biboptions{numbers,sort&compress}
\usepackage{amsmath,amssymb,graphicx,booktabs,longtable,array}
\usepackage[T1]{fontenc}
\usepackage[utf8]{inputenc}
\usepackage{lmodern}
\usepackage{microtype}
\usepackage{xurl}
\usepackage[hidelinks]{hyperref}
\journal{Physica A: Statistical Mechanics and its Applications}

\begin{document}
\begin{frontmatter}
\title{Amplifying or Redistributing Coupling: Synchronization in Sparse Adaptive Oscillator Networks}
\author[tm]{Yousef Jamali\corref{cor1}}
\cortext[cor1]{Corresponding author. E-mail: \href{mailto:y.jamali@modares.ac.ir}{y.jamali@modares.ac.ir}}
\address[tm]{Complex Systems Laboratory, Department of Applied Mathematics, School of Mathematical Sciences, Tarbiat Modares University, Tehran, Iran}
\begin{abstract}
Synchronization---different units falling into a common rhythm---matters in systems ranging from electrical generators to populations of neurons. But what happens when the connections that coordinate their timing can also change? If two units become more closely aligned and their connection strengthens, that extra influence can either increase the total coupling received by a unit or take a larger share of a fixed total. We examine these alternatives in a mathematical network of oscillators, each representing a repeating cycle, connected by sparse directed links. We compare constant connections with two adaptive rules that favor phase-aligned neighbors: unconstrained Hebbian coupling and homeostatic normalization, which keeps the total incoming effective coupling of each oscillator constant. Exact relationships establish this conservation and show that the mean Hebbian weight relaxes toward a value set by alignment along the network links. In matched numerical experiments, unconstrained adaptation reaches a given level of collective synchrony at weaker imposed coupling than fixed connections, whereas the normalized network requires stronger coupling within the same observation window. Longer observations reveal distinct individual trajectories, including increasing, sustained, and declining coherence. The central lesson is that the collective consequences of adaptation depend not only on which links are strengthened, but also on whether their strengthening adds influence or redistributes it.
\end{abstract}
\end{frontmatter}
\section{Introduction}
Keeping time together is important in systems that appear very different. Generators in an alternating-current power network must maintain compatible electrical rhythms, while interacting populations of neurons can show coordinated oscillatory activity. Both illustrate a broad scientific problem: how can units with different preferred rates nevertheless develop a shared pattern of timing? In the simplest mathematical description, each unit is an oscillator---something that repeatedly moves through a cycle---and its phase tells us where it is in that cycle. Synchronization occurs when the phases become more closely coordinated. The classical Kuramoto framework captures how interactions can produce such collective timing \cite{ref01,ref02,ref03}.

Who influences whom is just as important as how strongly they interact. A generator, a cell, or an abstract oscillator does not necessarily respond equally to every other unit. In a network model, each oscillator is a node; a directed link indicates that one node influences another; and the link weight measures how strongly it affects the receiver's timing. The architecture of these connections can shape collective coherence, and phase reduction explains when a more complicated oscillatory system can be represented by phase variables \cite{ref04,ref05,ref06}. Often, however, the connection strengths are not permanent. Activity-dependent changes in neural connections offer one motivation for studying adaptive networks, in which evolving states modify links and the modified links feed back into the states \cite{ref07,ref08}.

Consider a simple picture: one oscillator receives signals from eight neighbors. Two neighbors begin to move almost in step with it, while the other six remain less closely aligned. Suppose a learning rule favors the two well-aligned links. There are two fundamentally different ways to implement that preference. The two links may grow without reducing the others, increasing the total influence reaching the oscillator. Alternatively, their share may increase while the total incoming influence remains fixed, so other links must give up some of their share. The first process amplifies the input; the second redistributes it. Both reward alignment, but they do not impose the same collective dynamics. This distinction---additional coupling versus reallocation of existing coupling---is the question at the center of this paper.

Adaptive phase-oscillator models already show that the way connections change can transform collective behavior. Early work demonstrated learning through evolving coupling weights and enhanced synchronization in adaptive networks \cite{ref09,ref10}. Models motivated by synaptic plasticity revealed multistability and effects of communication delays \cite{ref11,ref12}. Subsequent studies examined the stability of Hebbian states, bifurcations, inertia, and the differences between reinforcing and weakening phase-dependent rules \cite{ref13,ref14,ref15,ref16}. The general message is clear: feedback between timing and connection strength can matter greatly. The more specific issue here is whether a gain in influence is allowed to increase a node's total incoming coupling.

Not all restrictions on adaptive connections mean the same thing. Competition among links can reorganize a network across different scales \cite{ref17}, and limits on available resources can change the conditions for synchronization \cite{ref18}. Related ideas arise in models of adaptive connections between spiking neurons \cite{ref19}. A restriction on the overall network, a resource constraint, and a rule that fixes every node's total incoming effective weight are nevertheless different mathematical assumptions. To understand their consequences, the meaning of the constraint must be specified rather than hidden inside the general term ``homeostasis.''

There are important precedents for both conservation and adaptation. Adaptive Kuramoto models with conserved incoming coupling have been analyzed mathematically \cite{ref20}, while other work has kept overall network coupling constant \cite{ref21}. Continuum theories of adaptive oscillators and frameworks with multiple adaptation time scales have further clarified their possible dynamics \cite{ref22,ref23}. Thus, neither phase-dependent learning nor conserved coupling is introduced here as a new idea. Our aim is a direct comparison in which the phase-learning preference is shared, but only one adaptive rule conserves the incoming effective coupling at each node.

A second complication is time. Even if the phases appear relatively well aligned, the connections that influence them may still be changing. Analytical results on conservation and rapid learning provide useful limits \cite{ref24}, while studies of synchronization and cluster stability show that adaptive coupling supports several organized states \cite{ref25,ref26}. Heterogeneous learning rules and differing adaptation time scales broaden this behavior further \cite{ref27,ref28,ref29,ref30}. It is therefore important to separate an exact property of the equations from what a numerical network happens to display after a particular observation time.

Indeed, the collective outcome of learning is not always that every oscillator falls into the same rhythm. Adaptive networks may organize into chimera states, clusters of different frequencies, or solitary oscillators \cite{ref31,ref32,ref33,ref34}. Multistability in star networks and transient groups in other network architectures show how much the outcome can depend on topology and initial conditions \cite{ref35,ref36}. General work on adaptive cluster patterns and coupled layers reinforces that lesson \cite{ref37,ref38}. For this reason, a population-average measure of synchronization is informative, but it cannot replace an examination of individual network trajectories.

Other forms of feedback provide a useful contrast. Local measures of coherence and adaptation between network layers can produce abrupt synchronization changes \cite{ref39,ref40,ref41}; adaptive interactions can also destabilize synchrony or generate unusual collective states \cite{ref42,ref43}. Higher-order interactions and multilayer architectures introduce additional mechanisms \cite{ref44,ref45}. These studies underscore the range of behaviors available to adaptive networks, but they should not be read as tests of the specific node-by-node incoming-weight budget considered here.

Recent research has continued to broaden the landscape through partial adaptation, external driving, and joint adjustment of natural frequencies and connection weights \cite{ref46,ref47,ref48}. Mean-field descriptions of synchronization-driven adaptation \cite{ref49} and reviews of higher-order adaptive interactions \cite{ref50} provide complementary perspectives. Together with the earlier literature, they motivate a controlled question rather than a claim of universal behavior: if phase alignment is rewarded in the same way, what changes when the total effective incoming strength is conserved instead of free to vary?

We address this question using a deliberately minimal model. Each unit has its own natural oscillation rate and receives weighted signals from a small, fixed number of neighbors in a directed network. We compare three cases on exactly matched graphs, frequencies, and starting phases: fixed links (Static); phase-dependent Hebbian links whose incoming total can change (Hebbian); and links with the same raw Hebbian preference but normalized so that each node always receives the same total effective weight (Homeostatic). We then increase the imposed coupling scale and measure how closely the phases line up during a specified observation window. The model is not a simulation of a particular brain circuit or electrical grid; those are motivations for asking how adaptive influence is allocated.

The comparison gives a clear answer at the scale actually measured. Analytically, normalization preserves each node's total incoming effective strength exactly, while unconstrained Hebbian weights relax toward a target determined by phase alignment along the links. Numerically, Hebbian networks attain the same chosen level of global coherence at lower imposed coupling than networks with fixed weights; normalized networks require higher coupling over the observation window. Following individual Hebbian networks for longer also reveals that their coherence need not evolve in the same direction, even when the average trend rises. The broader point is simple: a rule saying which neighbors deserve greater influence is incomplete without a rule saying where that influence comes from.

\section{Model and numerical methods}
\subsection{Sparse directed network}
We study N phase oscillators on directed in-regular graphs, so every node receives exactly \(k_0\) edges from distinct nodes and self-loops are excluded. The binary adjacency matrix Cij therefore obeys \(\sum_j C_{ij}=k_0\) for each i. Unless noted otherwise, \(k_0\)=8. Natural frequencies are drawn independently from a Gaussian distribution with standard deviation \(\sigma_\omega\)=0.5, then centered by subtracting the sample mean. For each numerical realization, the graph, natural frequencies, and initial phases are held fixed across all three coupling rules.

\subsection{Phase and weight dynamics}
The oscillators evolve through their weighted phase differences, while the raw edge weights relax toward a phase-dependent target:

\begin{equation}
\frac{d\theta_i}{dt}=\omega_i+\frac{K}{k_0}\sum_j C_{ij}w_{ij}\sin(\theta_j-\theta_i),
\end{equation}
\begin{equation}
\tau\frac{du_{ij}}{dt}=1+a\cos(\theta_j-\theta_i)-u_{ij},\qquad C_{ij}=1.
\end{equation}
Here K sets the overall coupling scale, \(\tau\) is the weight-adaptation time scale, and \(0\leq a<1\) controls the cosine-Hebbian preference for aligned phases. We compare three rules: (i) Static, \(w_{ij}\)=1; (ii) Hebbian, \(w_{ij}\)=\(u_{ij}\); and (iii) Homeostatic, \(w_{ij}\)=\(k_0u_{ij}/(\sum_lC_{il}u_{il})\). Both adaptive models use the same raw-weight relaxation equation. Only the Homeostatic model rescales the effective weights entering the phase dynamics, imposing a fixed total incoming effective weight at every integration stage.

\subsection{Order parameters and finite-window crossing}
We measure population-wide phase coherence with \(R_1(t)=\left|N^{-1}\sum_j\exp(i\theta_j)\right|\). To distinguish this global measure from the alignment of directly connected oscillators, we also track \(C_E(t)=(Nk_0)^{-1}\sum_{ij}C_{ij}\cos(\theta_j-\theta_i)\), the mean cosine of phase differences along directed edges. The second-harmonic order parameter \(R_2\) is retained as a supplementary descriptor; it is not used to classify a transition. Continuum approaches such as the Ott--Antonsen reduction require additional assumptions and are not used to estimate these finite-network observables \cite{ref51}.

To compare the curves on a common scale, we use \(K_{50}^{(T)}\): the coupling obtained by linear interpolation where the ensemble-mean, time-averaged \(R_1\) reaches 0.5 after a specified burn-in and over an observation window T. The symbol denotes a protocol-dependent crossing, not a thermodynamic critical point. For paired differences between coupling rules, 95\% bootstrap percentile intervals were computed from 2,000 resamples of matched realization identifiers (random seed 12345).

\subsection{Numerical protocols}
The primary matched scan covered N=200, 400, 800, and 1600, with 20 independent realizations at each size. The coupling grid was K=0.70, 0.775, 0.85, 0.925, 1.00, 1.075, 1.15, 1.225, and 1.30. We used \(k_0\)=8, \(\sigma_\omega\)=0.5, a=0.85, \(\tau\)=1, and \(\Delta t\)=0.05, with a burn-in of 40 and an observation window of 30 time units; \(R_1\) was sampled at one-unit intervals. Within a realization, graph, frequencies, and initial phases were matched across models and K values. In total, the scan comprised 4\(\times\)20\(\times\)3\(\times\)9=2160 model--realization--parameter runs.

To examine slower evolution, a targeted protocol used N=400 and 1600, six paired realizations per size, burn-in=100, observation=300, and \(\Delta t\)=0.05. The mechanistic analysis considered the six Hebbian realizations at N=1600, K=0.775, \(\tau\)=1, and a=0.85. Global coherence and edge/weight diagnostics were sampled every two time units and summarized in five successive 60-unit blocks. We reused the seeds, graphs, frequency vectors, and initial phases from the earlier 300-unit validation; the reproduced \(R_1\) traces agreed exactly with the archived traces.

The phase and raw-weight variables were advanced together using classical fourth-order Runge--Kutta integration. At every intermediate RK stage, both sets of variables were updated and both derivatives evaluated at that stage's state. The Supplementary Information gives implementation and numerical-verification details.

\section{Analytical properties}
\subsection{Normalization fixes incoming effective strength exactly}
Homeostatic normalization changes the allocation of influence but not its total: for each node with a positive sum of incoming raw weights, the effective weights obey

\begin{equation}
\sum_j C_{ij}w_{ij}^{\mathrm{HC}}=\sum_j C_{ij}\frac{k_0u_{ij}}{\sum_l C_{il}u_{il}}=k_0.
\end{equation}
The equality holds at every instant, independently of the oscillator phases. It is also well-defined throughout the simulated dynamics. Because \(0\leq a<1\), the target \(1+a\cos(\Delta_{ij})\) is always at least \(1-a>0\); positive initial raw weights remain positive under linear relaxation, keeping the normalization denominator nonzero.

\subsection{A fixed input budget favors locally aligned edges}
To see which edges gain influence within that fixed sum, consider quasi-static raw weights \(u_{ij}^*=1+a\cos(\Delta_{ij})\) and define the mean alignment among the neighbors entering node i as \(\bar c_i\)=\(k_0^{-1}\sum_l C_{il}\cos(\Delta_{il})\). Then

\begin{equation}
w_{ij}^{\mathrm{HC}}=\frac{1+a\cos(\Delta_{ij})}{1+a\bar c_i}.
\end{equation}
Expanding the denominator for small a gives

\begin{equation}
w_{ij}^{\mathrm{HC}}=1+a\left[\cos(\Delta_{ij})-\bar c_i\right]+O(a^2).
\end{equation}
The correction is determined by alignment relative to the local mean, rather than by alignment alone. An incoming edge whose phases are more aligned than the neighborhood average gains effective weight, while a less-aligned edge loses weight; the first-order changes sum to zero. This expression explains the redistribution mechanism at small a. It is not used as a quantitative approximation for the simulations at a=0.85.

\subsection{Fast Hebbian adaptation changes the shape of phase coupling}
If unconstrained raw weights relax infinitely rapidly (the singular limit \(\tau\to0\)), they approach \(u_{ij}\)=\(1+a\cos(\Delta_{ij})\), so the weighted phase interaction becomes

\begin{equation}
u_{ij}^*\sin(\Delta_{ij})=\sin(\Delta_{ij})+\frac{a}{2}\sin(2\Delta_{ij}).
\end{equation}
The ordinary sine interaction is thus accompanied by a second harmonic. This conclusion concerns the singular fast-adaptation limit; the simulations below retain the full coupled phase--weight dynamics at finite \(\tau\).

\subsection{Edge alignment sets the target for mean Hebbian weight}
A complementary identity holds at any adaptation time. Let \(E=Nk_0\) denote the number of directed edges, \(\bar u(t)=E^{-1}\sum_{ij}C_{ij}u_{ij}(t)\) the mean raw weight, and \(C_E(t)=E^{-1}\sum_{ij}C_{ij}\cos(\theta_j-\theta_i)\) the edge-alignment observable. Averaging the Hebbian weight equation gives

\begin{equation}
\tau\frac{d\bar u}{dt}=1+aC_E(t)-\bar u(t).
\end{equation}
This equality is exact: the mean raw weight relaxes toward \(1+aC_E(t)\), a target determined by phase differences along actual network edges. Rapid changes in edge alignment are filtered by the adaptation time \(\tau\), whereas slowly varying alignment is followed more closely. The relation is useful for interpreting weight and coherence trajectories without treating their observed association as proof of a separate causal pathway.

\subsection{Positive weights and the zero-plasticity case}
For \(0\leq a<1\), the target \(U_{ij}(t)=1+a\cos(\Delta_{ij}(t))\) lies in [\(1-a\),1+a]. The solution of \(\tau\dot u=U(t)-u\) is an exponentially weighted combination of the initial weight and the positive target history, so positive initial weights remain positive. At a=0, the raw weights relax toward unity; if they start at unity, all three rules coincide with the static phase equation throughout the evolution. Otherwise, that coincidence is reached only after the weight transient has decayed.

\section{Results}
\subsection{Amplification and redistribution shift the observed coherence response}
The three coupling rules separate systematically when global coherence is plotted against imposed coupling (Fig. 1). At the same finite observation time, unconstrained Hebbian networks attain \(R_1\)=0.5 at lower K than static networks, whereas the homeostatically normalized networks attain that level only at higher K. The ordering is the same at all four network sizes. Because each comparison uses matched graphs, frequencies, and initial phases across 20 realizations, the separation is attributable to the different coupling rules within the specified protocol, rather than to independently drawn network samples.

\begin{figure}[htbp]
\centering
\includegraphics[width=0.95\linewidth]{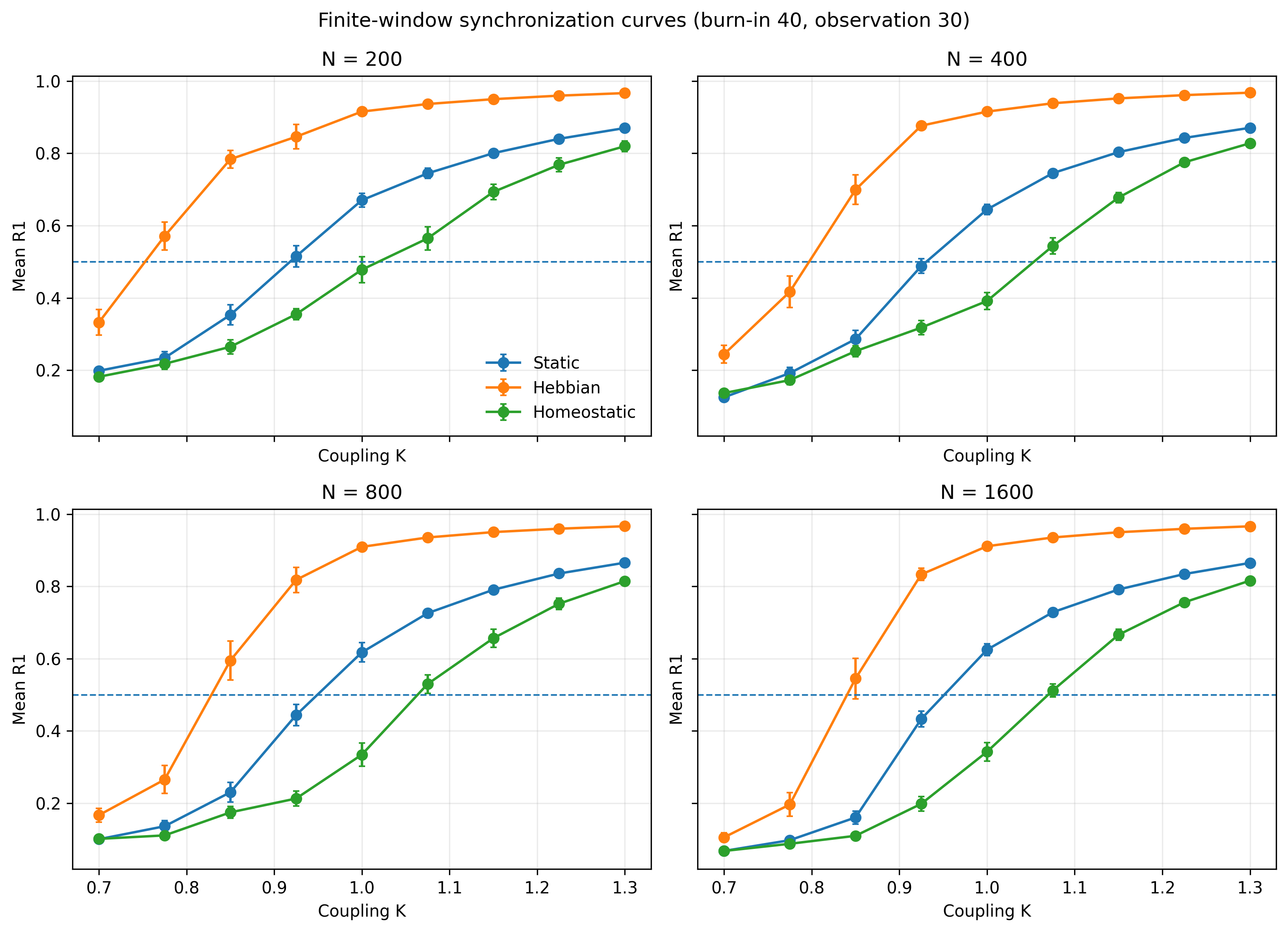}
\caption{Ensemble-mean finite-window \(R_1(K)\) curves. Error bars show the standard error across 20 realizations. The horizontal dashed line marks \(R_1\)=0.5.}
\label{fig:1}
\end{figure}
The interpolated \(K_{50}^{(T)}\) values support this ordering. For N=200, 400, 800, and 1600, the Static crossings are 0.918, 0.931, 0.949, and 0.951; the Hebbian crossings are 0.753, 0.797, 0.828, and 0.840; and the Homeostatic crossings are 1.019, 1.053, 1.064, and 1.070. Table 1 and Fig. 2 report the corresponding paired shifts and 95\% bootstrap intervals. These quantities compare finite-window responses; they do not by themselves locate asymptotic phase transitions.

\begin{table}[htbp]
\centering
\caption{Bootstrap estimates and 95\% intervals for finite-window $K_{50}$ shifts relative to Static.}
\label{tab:k50}
\small
\begin{tabular}{rlrrr}
\toprule
$N$ & Contrast & $\Delta K_{50}$ & 95\% CI low & 95\% CI high\\
\midrule
200 & Hebbian--Static & -0.165 & -0.182 & -0.148\\
200 & Homeostatic--Static & 0.101 & 0.079 & 0.135\\
400 & Hebbian--Static & -0.134 & -0.152 & -0.118\\
400 & Homeostatic--Static & 0.123 & 0.112 & 0.135\\
800 & Hebbian--Static & -0.121 & -0.138 & -0.101\\
800 & Homeostatic--Static & 0.114 & 0.102 & 0.129\\
1600 & Hebbian--Static & -0.111 & -0.124 & -0.093\\
1600 & Homeostatic--Static & 0.118 & 0.112 & 0.126\\
\bottomrule
\end{tabular}
\end{table}
\begin{figure}[htbp]
\centering
\includegraphics[width=0.95\linewidth]{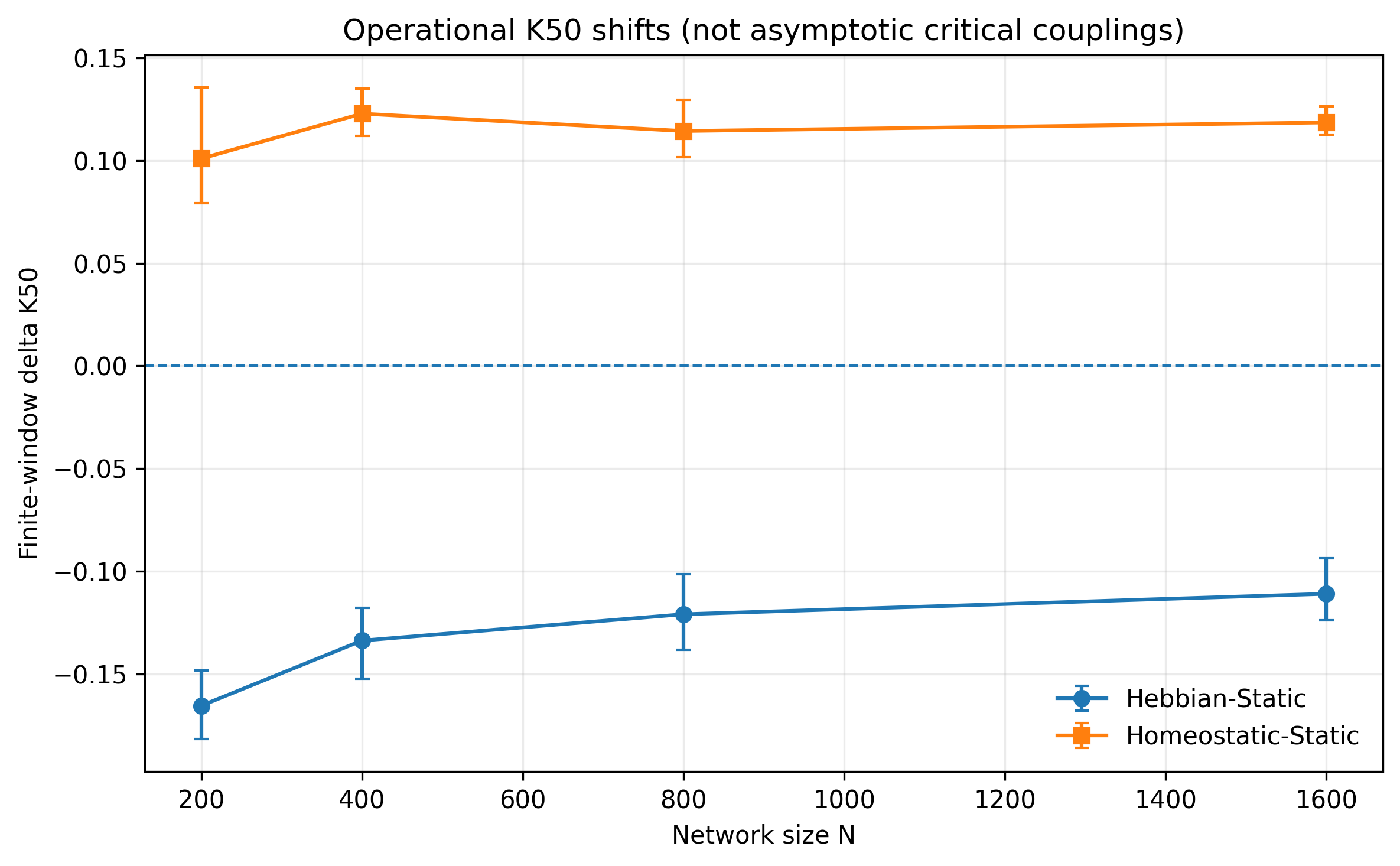}
\caption{Paired finite-window \(K_{50}\) shifts relative to Static, with 95\% bootstrap intervals.}
\label{fig:2}
\end{figure}
\subsection{Near the crossing, the network can still be evolving}
A common crossing level does not imply that the underlying trajectories have stopped changing. Near \(K_{50}^{(T)}\), the difference between early and late halves of the observation window remains most pronounced for Hebbian coupling (Fig. 3). The median absolute half-window differences for Hebbian are 0.0808, 0.0720, 0.0921, and 0.0517 at N=200, 400, 800, and 1600. The corresponding values are 0.0422, 0.0207, 0.0107, and 0.0133 for Static, and 0.0260, 0.0477, 0.0342, and 0.0317 for Homeostatic.

\begin{figure}[htbp]
\centering
\includegraphics[width=0.95\linewidth]{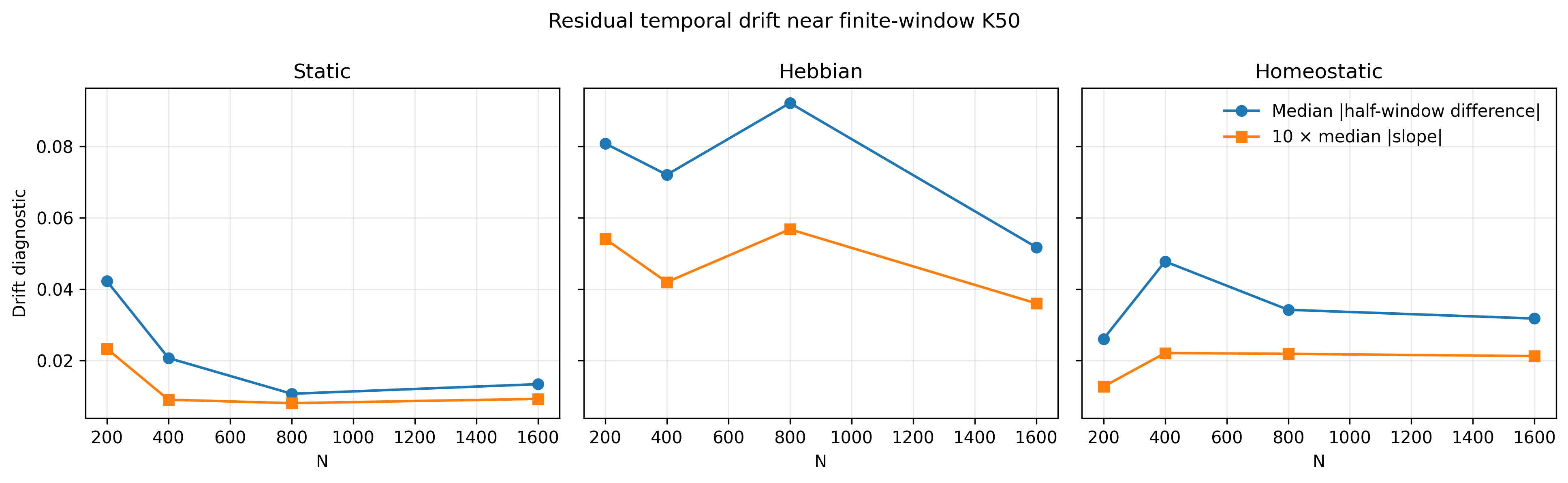}
\caption{Residual temporal-drift diagnostics near the finite-window crossing. The plotted slope is rescaled for visual comparison; the underlying values are provided in the Supplementary Information.}
\label{fig:3}
\end{figure}
The half-window differences and sampled slopes indicate that relaxation can remain visible precisely where a crossing is being measured. This observation motivates the longer-window analysis below: the crossing curves describe how the models respond over the chosen period, while the trajectories reveal how that response continues to unfold.

\subsection{Longer observation reveals different Hebbian routes to coherence}
Extending the Hebbian observations from 30 to 300 time units does not reveal one uniform route to synchronization. For six N=1600 realizations at K=0.775, the ensemble-mean \(R_1\) increases across five successive 60-unit blocks, from 0.428 in the first to 0.590 in the fifth. Yet the mean conceals substantial heterogeneity: four realizations gain coherence, one stays at a highly coherent level, and one loses coherence. Thus, even with a common model and coupling strength, different realizations can move in different directions over the same observation period.

\subsection{Phase alignment and edge weight evolve together}
The longer trajectories also show how the adaptive couplings change as coherence evolves. Across the same five blocks, the mean raw edge weight rises from 1.265 to 1.382, while the edge-alignment observable \(C_E\) increases from 0.314 to 0.450 (Fig. 4). The associated target \(1+aC_E\) changes from 1.267 to 1.383, closely tracking the observed weight, just as the exact mean-weight relaxation identity predicts. The measured rise in global coherence is therefore accompanied by a rise in alignment specifically along the edges that carry coupling.

\begin{figure}[htbp]
\centering
\includegraphics[width=0.95\linewidth]{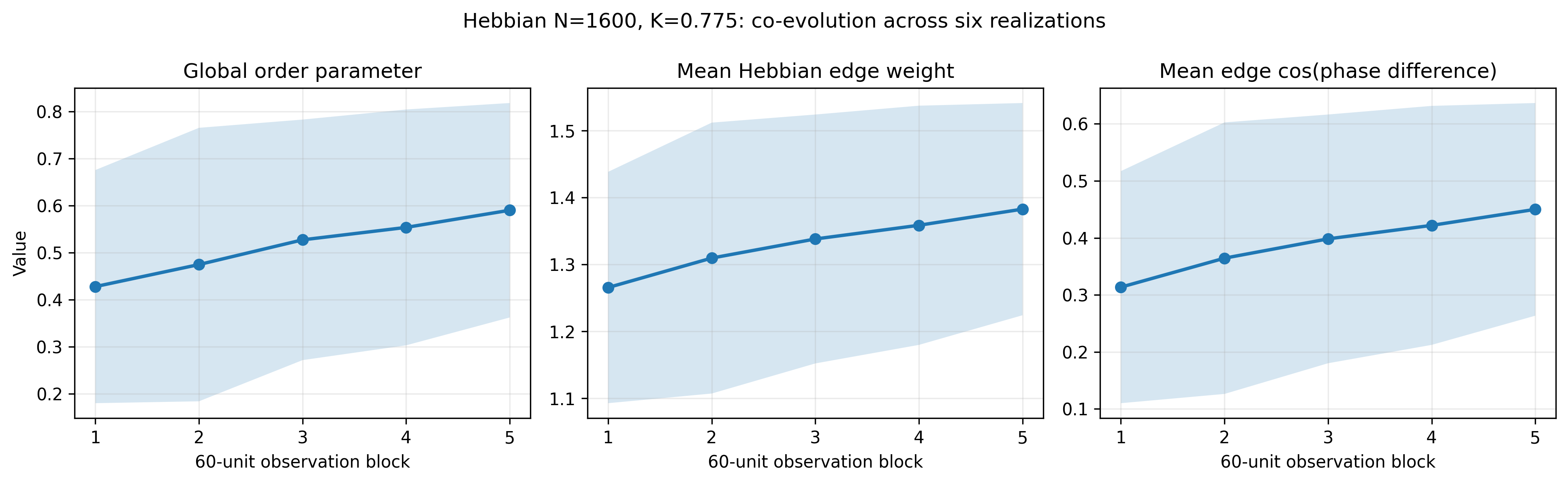}
\caption{Block-level co-evolution of global coherence, mean raw edge weight, and edge phase coherence in the six N=1600 Hebbian realizations.}
\label{fig:4}
\end{figure}
\begin{figure}[htbp]
\centering
\includegraphics[width=0.95\linewidth]{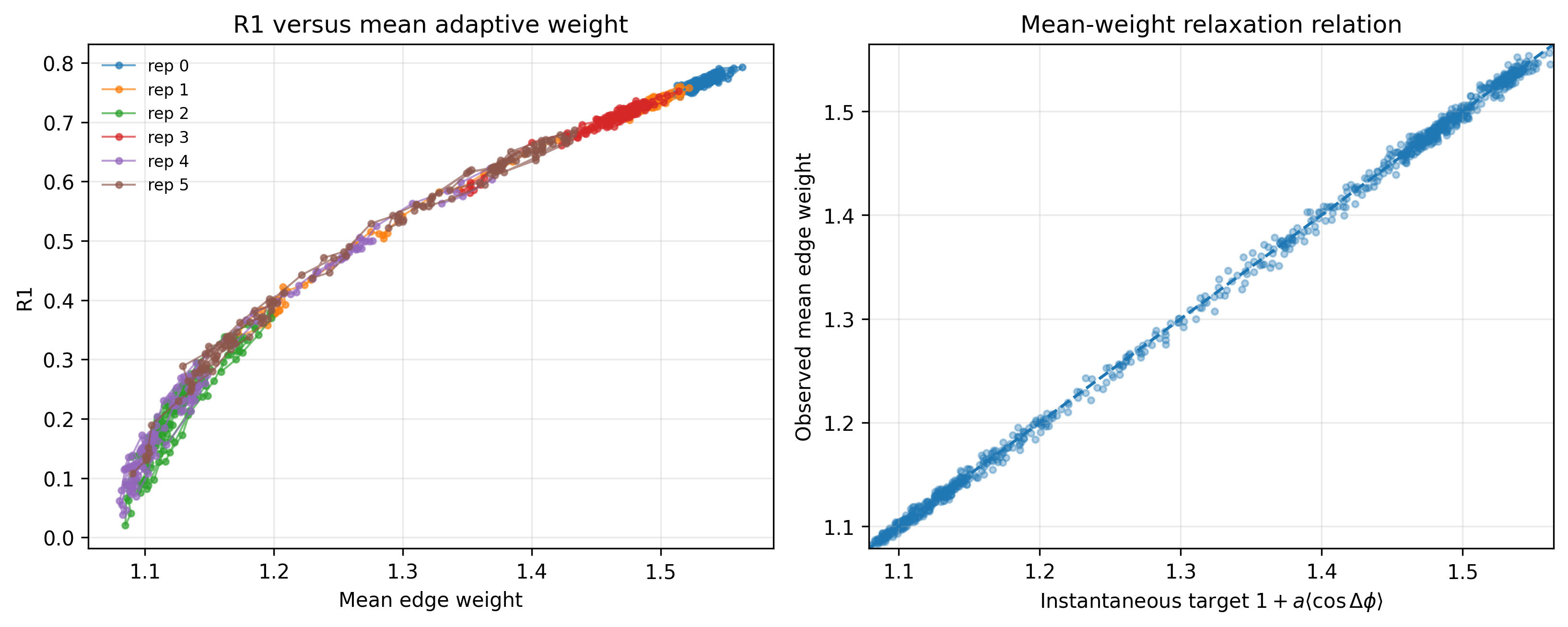}
\caption{Within-realization association between \(R_1\) and mean edge weight, and comparison of observed mean weight with the instantaneous target \(1+aC_E\).}
\label{fig:5}
\end{figure}
Within individual realizations, the mean Pearson correlations are 0.955 between \(R_1\) and mean raw weight, 0.982 between \(R_1\) and \(C_E\), and 0.973 between mean raw weight and \(C_E\) (Fig. 5). These autocorrelated time series make the correlations descriptive rather than independent statistical tests. More importantly, the exact relaxation equation explains why the mean weight follows edge alignment. The realization in which all three measures decline shows that the relationship also accompanies decreasing coherence; it is not simply a consequence of averaging several increasing traces.

\section{Discussion}
\subsection{A shared learning rule can have two different collective effects}
The central distinction is not whether phase alignment strengthens a connection; both adaptive models favor aligned neighbors in their raw-weight dynamics. It is whether this preference can change the total effective input received by each oscillator. Without normalization, several incoming connections may grow together and amplify the node's net coupling. With normalization, the same preferences compete for a fixed sum. The exact conservation identity and the small-a expansion make this difference explicit, separating a rule for learning edge preferences from a rule for allocating coupling strength.

A second analytical perspective explains why these alternatives matter dynamically. In the fast-adaptation limit, the unconstrained Hebbian interaction contains a second-harmonic contribution, changing the functional form of coupling. At finite adaptation times, the mean raw weight does not respond instantaneously: it relaxes toward a target set by edge-specific phase alignment. The numerical trajectories exhibit this tracking. Taken together, the theory and observations show how collective coherence, edge alignment, and adaptive weights become linked while leaving room for distinct realization-level outcomes.

\subsection{A finite-time synchronization curve is a snapshot of an evolving network}
The ordered separation of the crossing curves is a robust feature of the matched finite-window comparison: the unconstrained Hebbian model reaches the chosen coherence level at weaker imposed coupling, while the normalized model reaches it at stronger coupling. The drift diagnostics add an essential temporal interpretation. The measured curve is not only a response to K; it also reflects how far phase and weight dynamics have evolved before and during observation. This matters whenever adaptive networks are compared over a fixed experimental or numerical time horizon, because apparently sharp differences in coherence can coexist with slow internal adjustment.

\subsection{Position within adaptive-network research and broader implications}
Earlier work demonstrated that Hebbian plasticity can support synchronization, clusters, and multistability \cite{ref09,ref11,ref13,ref14}, while conservation and resource constraints can change network behavior \cite{ref18,ref21,ref23,ref24}. The present study does not introduce either idea in isolation. Its contribution is to place a specific nodewise effective in-strength constraint beside an unconstrained rule with the same raw adaptation law, and to connect that controlled comparison to exact weight identities and matched finite-time responses. This also prevents treating distinct forms of 'homeostasis' as if they imposed the same physical restriction.

The realization-dependent dynamics fit a wider picture in which topology, clusters, and transients shape adaptive collective behavior \cite{ref29,ref30,ref32,ref33,ref37}. Related research connects adaptive neuronal-network ideas to power-grid dynamics \cite{ref52}. Such systems motivate broader questions about how adaptive interactions are budgeted, but the present calculations do not model neurons or power grids directly. What they offer is a conceptual diagnostic: before interpreting a change in synchronization, ask whether adaptation changes total effective coupling, redistributes it, or does both.

The numerical results are likewise distinct from analyses of explosive transitions, higher-order network interactions, or asymptotic criticality in other adaptive models \cite{ref39,ref45,ref50}. Here, their value lies in the controlled sparse-network comparison and the transparent relation between global coherence, local edge alignment, and mean-weight relaxation. This perspective suggests a way to organize future comparisons among adaptation rules without presuming that all observed crossing shifts have the same physical origin.

\subsection{Scope and questions for further study}
The reported \(K_{50}^{(T)}\) values depend on finite burn-in and observation windows; the near-crossing drift makes an asymptotic critical-point interpretation unwarranted. The detailed weight--coherence analysis covers six Hebbian trajectories at N=1600 and K=0.775. The small-a and \(\tau\to0\) expressions are analytical limits, whereas the simulations use a=0.85 and \(\tau\)=1, and all numerical comparisons use sparse directed in-regular graphs. Natural next tests would separate weight amplification from weight redistribution under matched effective-input strength, and vary observation time, degree, and topology to determine which aspects of the finite-time ordering persist.

\section{Conclusions}
How a connection adapts is only part of the story; what matters equally is how the network pays for that adaptation. In sparse oscillator networks, a phase-dependent preference for aligned neighbors can either amplify incoming coupling or redistribute a conserved amount. Exact identities make the distinction visible at the level of individual nodes and edges, while matched simulations show that it shapes the observed development of global coherence. Longer trajectories reveal why the collective response cannot be understood from a crossing curve alone: phase alignment and weights continue to evolve, and different realizations follow different paths. The broader lesson is that adaptation rules should be interpreted together with the constraints under which they operate. That perspective can help distinguish seemingly similar forms of collective learning and synchronization across adaptive network models.

\section*{Declarations}
\subsection*{Funding}
This research received no specific grant from any funding agency in the public, commercial, or not-for-profit sectors.

\subsection*{Declaration of competing interest}
The author declares that he has no known competing financial interests or personal relationships that could have appeared to influence the work reported in this paper.

\subsection*{CRediT authorship contribution statement}
Yousef Jamali: Conceptualization, Methodology, Formal analysis, Software, Validation, Investigation, Data curation, Visualization, Writing -- original draft, Writing -- review \& editing, Project administration.

\subsection*{Acknowledgements}
None.

\subsection*{Data and code availability}
The simulation parameters and details of the analytical procedures are provided in the main text and Supplementary Information. The code used to generate the numerical results and figures is available from the corresponding author upon reasonable request.

\subsection*{Declaration of generative AI and AI-assisted technologies in the manuscript preparation process}
During the preparation of this manuscript, the author used OpenAI ChatGPT to edit and improve the language, structure, clarity, and readability of the text. The author critically reviewed and approved the final manuscript and takes full responsibility for its content.


\end{document}